\documentclass[a4paper, 10pt, conference]{ieeeconf}         % A4

\IEEEoverridecommandlockouts
\usepackage{graphics}
\usepackage{epsfig}
\usepackage{mathptmx}
\usepackage{times}
\usepackage{amsmath}
\usepackage{amssymb}
\usepackage{filecontents}
\usepackage{comment}
\usepackage{mathrsfs}
\usepackage{eucal}
\usepackage{cuted}
\usepackage{flushend}
\usepackage{float}
\let\labelindent\relax
\usepackage{subcaption}
\usepackage{nicefrac}
\usepackage{enumitem}
\usepackage{verbatim}
\usepackage{siunitx}
\usepackage{textcomp}
\usepackage{tikz}
\usepackage{amsfonts}
\usepackage{mathtools}

\usepackage{algorithmic}
\usepackage{graphicx}
\usepackage{xcolor}
\usepackage{booktabs}
\usepackage{url}
\usetikzlibrary{shapes,arrows}
\usepackage[europeanresistors,americaninductors,cuteinductors]{circuitikz}
\usepackage[normalem]{ulem}

\usepackage[style=ieee, isbn=false, doi=true, url=false]{biblatex}

\title{\LARGE \bf
Retrievable but Unencountered: The Missing Exposure Denominator \\ in Large Academic Ebook Collections}

\author{Jette Veenstra$^{1}$ and Mauricio Mu\~noz-Arias$^{2}$% <-this % stops a space
\thanks{$^{1}$Jette Veenstra is with the University Library, University of Groningen, Groningen, The Netherlands.
        {\tt\small jette.veenstra@rug.nl}}%
\thanks{$^{2}$Mauricio Mu\~noz-Arias is with the Faculty of Science and Engineering, University of Groningen, Groningen, The Netherlands.
        {\tt\small m.munoz.arias@rug.nl}}%
}
\begin{document}

\maketitle
\thispagestyle{empty}
\pagestyle{empty}

%%%%%%%%%%%%%%%%%%%%%%%%%%%%%%%%%%%%%%%%%%%%%%%%%%%%%%%%%%%%%%%%%%%%%%%%%%%%%%%%
\begin{abstract}

Academic libraries hold ebook collections so large that no reader can inspect more than a negligible fraction of the catalogue.
Assessment of those collections rests almost entirely on usage telemetry.
A title that generates no recorded interaction is treated as a title nobody wanted.
That inference is unavailable on the present evidence.
We separate two constructs the literature conflates: retrievability, the probability that a system returns an item to a query that already specifies it, and encounterability, the probability that an item enters a reader's attention without prior title-level intent.
Following a stated protocol over work published from 2018 onward, we find three things.
Large zero-use populations are robustly documented.
Usage depends heavily on the measuring instrument.
And no standard library metric records an item impression.
The consequence is an identification problem.
Recorded non-use mixes non-exposure, unnoticed exposure, rejection, and off-channel use, and no study located under our protocol decomposes it.
Recommender-system research already names this problem: there, feedback is missing not at random.
We close with four measurement directions that would narrow the gap without closing it.

\end{abstract}

%%%%%%%%%%%%%%%%%%%%%%%%%%%%%%%%%%%%%%%%%%%%%%%%%%%%%%%%%%%%%%%%%%%%%%%%%%%%%%%%
\section{INTRODUCTION}

A library that has moved its monographs online can say with confidence that every title it owns is indexed, resolvable, and retrievable.
Supply an author, a title, an identifier, or a sharp enough subject query, and the system returns the book.
That is a real achievement, and it is routinely offered as proof that the collection is discoverable.
The claim does not survive a change of question.
Retrieval tests use queries that already presuppose the target, whereas the operationally interesting question for a collection of a million titles runs the other way: how does a book the reader has never heard of become a book the reader knows about?
The distinction is old in information behaviour.
Bates described real search as berrypicking across a query that shifts as the searcher learns, with browsing a legitimate strategy rather than a degenerate form of search~\cite{Bates1989}, and Marchionini split lookup from the exploratory modes of learning and investigating~\cite{Marchionini2006}.
It is also consequential.
A title with no recorded use is a candidate for non-renewal, for removal from future purchase profiles, and for demotion in ranked results, and every one of those decisions reads a zero as a revealed preference.
But a zero may instead record an item nobody was ever shown.
Then the decision is not evidence-based.
It is a feedback loop.

Three bodies of work bear on this question, and they have not been brought together.
The first measures ebook use at title level and documents its unevenness: large zero-use populations appear under conventional acquisition~\cite{Fry2018}, under evidence-based acquisition~\cite{Yilmaz2022,Strothmann2020,Tran2021}, and across vendors over time~\cite{Tracy2019}, while methodological work shows that what counts as use is itself instrument-dependent~\cite{Hughes2020,Snijder2021,Echeverria2023,COUNTER51} and that acquisition-model outcomes observe demand only after a title has already reached a patron~\cite{Downey2020,Tyler2019}.
The second theorises the missing step.
Erdelez and Makri propose information encountering as the preferred construct and give it a temporal model~\cite{Erdelez2020}; Bj\"orneborn's affordance account explains why the move to electronic form is not neutral, physical environments holding a primacy in sensoriability and digital environments in traversability~\cite{Bjorneborn2017}; and empirical work has begun to model browsing itself~\cite{McKay2019,Makri2019,Zhang2022}, though a recent survey finds the field fragmented~\cite{Liu2022}.
The third defines exposure rigorously, but elsewhere.
Recommender-system research treats implicit feedback as missing not at random precisely because the system controls exposure~\cite{Saito2020,Castells2022}, and names the mechanisms by which visible items stay visible and new ones stay hidden~\cite{Abdollahpouri2021,Klimashevskaia2024,Panda2022,Diaz2020}.
The first body of work can see use and not exposure.
The second can characterise encounter and does not instrument it at item level.
The third instruments exposure on commercial platforms that are not libraries.
No standard library metric records an item impression, and no located study joins exposure to subsequent use on the same collection.

This paper does not close that gap.
It characterises it, and the contribution is fourfold.
First, we separate two constructs the literature conflates: \emph{retrievability}, the probability that a system returns an item given a query that specifies it, and \emph{encounterability}, the probability that an item enters a reader's field of attention during activity aimed at something else.
We then show that a recorded zero decomposes into four states, only one of which is a statement about the reader's preference.
Second, because several of our claims are negative existence claims, we state a review protocol and the limits of what it supports, so that those claims can be judged rather than taken on trust.
Third, we import the formal statement of the problem from recommender-system research and tabulate the dependent variables in current use against what each can and cannot detect, which locates every established library measure at or after the third stage of an availability-exposure-selection-use-citation ordering.
Fourth, we set out four measurement directions that would narrow the gap, and state in each case what would remain unresolved.
This is a problem paper, and we say so plainly.
We propose no solution.
The absence of one is our finding, not a gap in our treatment.

The remainder of the paper is organised as follows.
Section~II separates the two constructs and places them in the information-behaviour literature.
Section~III states the review protocol, without which the negative claims later on cannot be judged.
Sections~IV to~VI assemble the item-level evidence on zero-use, show why the instruments cannot observe exposure, and ask how thin the library's own exposure surface is.
Section~VII imports the formal statement of the problem from recommender-system research, and Section~VIII tabulates how the problem has been operationalised.
Section~IX sets out four measurement directions and says what each would leave unresolved.
Section~X records the limits of the evidence base and of our own protocol.
Section~XI concludes.

%%%%%%%%%%%%%%%%%%%%%%%%%%%%%%%%%%%%%%%%%%%%%%%%%%%%%%%%%%%%%%%%%%%%%%%%%%%%%%%%
\section{TWO CONSTRUCTS, ONE MEASUREMENT}

The idea is old.
Information seeking is not exhausted by matched retrieval.
Bates described real search as berrypicking: an evolving process across a query that shifts as the searcher learns, with browsing a legitimate strategy rather than a degenerate form of search~\cite{Bates1989}.
Marchionini split lookup from the exploratory modes of learning and investigating~\cite{Marchionini2006}.
That second mode is exactly the regime in which a reader has no title in mind.
Both papers predate 2018.
We cite them as lineage, not as evidence.

The current formulation is information encountering.
Erdelez and Makri argue that loose terminology has held cumulative research back~\cite{Erdelez2020}.
They propose information encountering as the preferred construct and give it a temporal model, one in which an encounter is an event inside a sequence rather than a synonym for acquisition.
Buchanan and Erdelez carry this into humanities practice, where encounter binds up with disciplinary rhythm and with later rediscovery and sharing~\cite{Buchanan2019}.
Liu et al.\ survey the field and find it fragmented across information behaviour and recommender systems~\cite{Liu2022}.
The two bodies of work rarely speak to each other.

The affordance account explains why going electronic is not neutral for encounter.
Bj\"orneborn names three environmental affordances for serendipity: diversifiability, traversability, and sensoriability~\cite{Bjorneborn2017}.
Physical environments, he argues, hold a primacy in the third; digital environments hold a primacy in the second.
An ebook collection gains enormously in traversability.
What it loses is the channel through which a shelf makes its neighbours perceptible without being asked.

Empirical work on browsing has begun to catch up.
It has not yet reached item level.
McKay et al.\ build a typology of library browsing and note that moving a collection online renders it invisible to readers who prefer to browse~\cite{McKay2019}.
Makri et al.\ name the split directly: digital environments are built for goal-directed seeking over loosely directed exploration, for searching over discovering~\cite{Makri2019}.
Zhang et al.\ give one of the few empirical models of online browsing, a four-part iterative cycle of choosing a collection, selecting candidates, inspecting them, and revising the candidate list~\cite{Zhang2022}.
Modern systems, they observe, are tuned for query-based search and neglect browsing.

These are qualitative studies of process.
None joins a browsing episode to a title-level record of what was later used.
The construct is well theorised.
It is unmeasured at the level where a library must act.

%%%%%%%%%%%%%%%%%%%%%%%%%%%%%%%%%%%%%%%%%%%%%%%%%%%%%%%%%%%%%%%%%%%%%%%%%%%%%%%%
\section{METHOD}

Several claims below are negative existence claims.
Such a claim is worth no more than the search behind it.
We therefore state the protocol in full.
This is a structured narrative review, not a systematic review in the sense of PRISMA (Preferred Reporting Items for Systematic Reviews and Meta-Analyses).
Section~X records what that costs.

\subsection{Scope and eligibility}
The evidence tier holds only work published from 2018 onward.
Older work enters for theoretical lineage alone.
It never supports an empirical claim, and we mark it as lineage at the point of citation~\cite{Bates1989,Marchionini2006,Bjorneborn2017}.
To enter the evidence tier, a source had to be a peer-reviewed journal article or a refereed conference paper.
It also had to do one of two things: report title-level data on an academic ebook collection, or make an explicit conceptual contribution to the encounter construct.
Grey literature, standards, and vendor documentation enter on narrower terms.
They document current practice.
They never evidence an effect, and we flag them wherever they appear~\cite{COUNTER51,Blankstein2022,OCLC2024,Clamon2019}.

\subsection{Sources and query families}
We searched the ACM Digital Library, arXiv, Emerald, MIT Press, SAGE, ScienceDirect, Springer, Taylor \& Francis, and Wiley, together with the journal platforms of the American Library Association (ALA) and the Association of College and Research Libraries (ACRL).
General web search supplied the grey and vendor material.
Five query families ran across them: item-level ebook usage and zero-use; acquisition-model outcomes under demand-driven and evidence-based acquisition; browsing, serendipity, and information encountering in born-digital collections; virtual-shelf, call-number-browse, and new-arrivals display effects; and exposure bias, popularity bias, cold start, and long-tail coverage in recommender and information-retrieval systems.
Representative strings include \emph{ebook zero-use item-level academic
library COUNTER}; \emph{virtual shelf call number browse discovery layer
evaluation}; \emph{new books carousel discovery interface increased
circulation}; and \emph{exposure metric library collection encounter
recommender}.

\subsection{Access levels}
We classified every retrieved source by how far we could read it: full text, abstract and metadata only, or secondary mention inside another source.
Claims run only as far as the access supports.
Where we had the abstract alone, we report what the abstract states and say so.
Where a result reached us only through another paper's summary, we do not use it as evidence at all.

\subsection{What the protocol does not support}
Screening was not double-blinded.
We computed no inter-rater agreement and kept no record counts by stage.
The review therefore cannot report a PRISMA flow diagram, and we present none.
Read every negative statement in this paper as \emph{not located under the protocol above}.
None of them demonstrates that no such work exists.
We report the absences anyway, for two reasons.
Absence of located evidence bears on the argument directly.
And the alternative, quietly substituting an adjacent study that does not in fact answer the question, would misrepresent the state of the field.

%%%%%%%%%%%%%%%%%%%%%%%%%%%%%%%%%%%%%%%%%%%%%%%%%%%%%%%%%%%%%%%%%%%%%%%%%%%%%%%%
\section{WHAT THE ITEM-LEVEL EVIDENCE ESTABLISHES}

The empirical picture is consistent across institutions, acquisition models, and countries.
Use is severely uneven.
In any finite window, a large share of available titles generates no recorded interaction at all.
Table~\ref{tab:zerouse} collects the studies for which we obtained item-level figures at full-text access.

\begin{table*}[!t]
\caption{Item-level ebook non-use in studies read at full text, 2018 onward}
\label{tab:zerouse}
\centering
\small
\begin{tabular}{@{}lrrp{9.0cm}@{}}
\toprule
\textbf{Study} & \textbf{Titles} & \textbf{No recorded use} & \textbf{Setting and qualification} \\
\midrule
Fry~\cite{Fry2018} & 73{,}148 & 88\% & One US academic library, 17-month window; 27\% of 39{,}342 print titles used over the same period. The author cautions that package-level circumstances complicated part of the ebook zero-use analysis, so 88\% is best read as an upper bound. \\[3pt]
Y{\i}lmaz \& \"Unal~\cite{Yilmaz2022} & 35{,}624 & 93\% & Evidence-based acquisition pool, Hacettepe University, Turkey; the entire pool was made accessible to users before any purchase decision. \\[3pt]
Strothmann \& Rupp-Serrano~\cite{Strothmann2020} & 3{,}781 & 39\% & Elsevier evidence-based collection, University of Oklahoma, US. \\[3pt]
Tran \& Guo~\cite{Tran2021} & 3{,}186 & 89\% & Evidence-based acquisition pilot in science and engineering, Stony Brook University, US; 12-month window, November 2019 to October 2020. The 361 used titles generated 6{,}810 downloads. \\
\bottomrule
\end{tabular}
\end{table*}

Fry compared print and electronic monographs acquired over overlapping periods at one institution~\cite{Fry2018}.
Of the ebooks acquired between 2008 and 2014, 12 percent recorded use in the observation window.
Print managed 27 percent over the same seventeen months.
Y{\i}lmaz and \"Unal report the most striking figure, and the most instructive one~\cite{Yilmaz2022}.
Of 35,624 titles in an evidence-based acquisition pool, 2,462 recorded use.
The other 33,162 did not.
The entire pool had been made accessible to users in advance of any purchase decision, which removes availability as an explanation and leaves the question of exposure entirely open.
Strothmann and Rupp-Serrano found 1,486 of 3,781 books in an Elsevier evidence-based collection with no recorded access~\cite{Strothmann2020}.
Neither an approval-plan simulation nor simulated librarian firm orders reproduced the subject distribution of actual patron access.

Tran and Guo run the same design in science and engineering~\cite{Tran2021}.
Of 3,186 titles made fully accessible before any purchase decision, 361 were used over twelve months, a rate of 11.33 percent.
Those 361 titles generated 6,810 downloads between them, so the use that did occur was concentrated rather than thin.
Their headline measure is percent used, and it illustrates the denominator problem exactly: it takes the accessible pool as its base, whereas discoverability needs the smaller and unobserved base of titles a reader had any opportunity to notice.
Tracy adds a temporal caution.
Analysing five years of usage in an interdisciplinary collection, he argues that the used-versus-unused binary loses diagnostic value as a collection matures, and documents that COUNTER (Counting Online Usage of NeTworked Electronic Resources) whole-book and section reports cannot be pooled~\cite{Tracy2019}.

What these studies establish is the existence and scale of the zero.
They do not establish its meaning.
A recorded zero is consistent with at least four distinct states:

\begin{enumerate}[leftmargin=*, itemsep=1pt]
\item the title was never rendered to any user;
\item it was rendered but not noticed;
\item it was noticed and not selected;
\item it was used through a channel the instrument did not observe.
\end{enumerate}

Only the third is a statement about the reader's preference.
The first two are properties of the system.
The fourth is a property of the instrument.
No study located under our protocol tells them apart.

%%%%%%%%%%%%%%%%%%%%%%%%%%%%%%%%%%%%%%%%%%%%%%%%%%%%%%%%%%%%%%%%%%%%%%%%%%%%%%%%
\section{WHY THE INSTRUMENTS CANNOT SEE EXPOSURE}

The limitation is structural.
It is not a matter of collecting too little data.

COUNTER title metrics begin after an interaction with content or with title metadata.
\texttt{Unique\_Title\_Requests} counts a title at most once per session, however many segments are requested.
That is a genuine gain in comparability: some platforms deliver whole books, others deliver chapters.
The Code recommends unique-title metrics for comparison across releases, because item-level counts can shift between R5 and R5.1~\cite{COUNTER51}.
But better denominator control is not an exposure denominator.
Nothing in the Title Reports records the occasions on which a book was rendered in a result list, a carousel, a shelf-browse strip, or a recommendation panel and not acted upon.

Two independent comparisons show that even post-exposure usage is instrument-relative.
Snijder measured one corpus of roughly 11,000 open-access books and chapters through two systems at once~\cite{Snijder2021}.
Google Analytics recorded about 3.6 million downloads.
COUNTER R5 recorded 1.6 million.
The two cannot be converted into one another mechanically, and country and title rankings diverge.
Echeverr\'ia and Bustamante compared COUNTER R5 against link-resolver activity across three platforms~\cite{Echeverria2023}.
Association was only moderate, with Spearman coefficients between 0.561 and 0.678 at $p < .01$, and the two capture different aspects of ebook interaction.
They also note that COUNTER R5 lacks coverage of zero-usage titles.
A purchased and never-accessed title can be absent from the report rather than present with a zero.
Tingle and Teeter reach the same conclusion from the practitioner side~\cite{Tingle2018}.
Assessing a promoted set of ebooks across eleven platforms, they report that zero-use data would have materially helped, and that few vendors supply it because the Code does not require it.
Two of their eleven platforms returned no title-level data at all.

Hughes provides the definitional argument.
Analysing four years of COUNTER R4 section-level reports at a research institution, he separates intensive measures such as page requests from extensive measures such as whether a title was used at all~\cite{Hughes2020}.
Recoding activity into a binary use representation removes most of the apparent volatility of page-level data.
He notes the exact print analogue: circulation records omit in-library consultation.
Change the event definition and you change what use means.
No available definition reaches back before the user acted.

The acquisition literature inherits the problem rather than solving it.
Downey and Zhang compare demand-driven acquisition across eight large libraries and show that withdrawing titles from the discovery pool too early alters observed long-term use~\cite{Downey2020}.
That is informative about the consequences of exposure while measuring only its downstream effect.
A demand-driven trigger is conditional, by construction, on a title having reached a patron.
The denominator of titles that reached patrons and were not triggered is never recorded.
Tyler et al.\ move the endpoint further downstream still, operationalising acquisition success as subsequent citation counts and naming this as a limit of their own design~\cite{Tyler2019}.
A citation shows that a work entered scholarly activity.
It says nothing about how the work first became known.

%%%%%%%%%%%%%%%%%%%%%%%%%%%%%%%%%%%%%%%%%%%%%%%%%%%%%%%%%%%%%%%%%%%%%%%%%%%%%%%%
\section{HOW THIN IS THE EXPOSURE SURFACE?}

Exposure is unmeasured.
Its plausible magnitude can still be bounded from adjacent evidence, and the bound is not reassuring.

The Ithaka S+R US Faculty Survey went to a national sample in autumn 2021 and drew 7,615 completed responses~\cite{Blankstein2022}.
Some 14 percent of faculty most often begin discovery of scholarly material at their institution's library website or online catalogue.
A third begin at a specific database, and 29 percent at Google Scholar.
Disciplinary divergence is pronounced: 36 percent of social scientists start at Google Scholar, against 12 percent of humanists.
So the library's own discovery surface is the entry point for a minority of the very population whose encounters it hopes to influence.
This is a self-report instrument, about scholarly literature in general rather than ebooks specifically.
We treat it as an order-of-magnitude bound, not a measurement.

The one browse affordance libraries built deliberately for the digital catalogue largely excludes the digital collection.
Virtual-shelf and call-number-browse features rebuild classification adjacency in the catalogue.
At least one major implementation restricts the feature to physical holdings: the documentation of the Online Computer Library Center (OCLC) for WorldCat Discovery states that \emph{Browse the Shelf only displays physical materials held by your institution}~\cite{OCLC2024}.
This is vendor documentation, not peer-reviewed evidence, and we cite it only as a record of current practice.
We located no study under our protocol measuring whether adding ebooks to call-number browse changes their encounter or use.

The physical evidence that position matters is comparatively strong.
Broadbent correlated 2.25 years of usage against shelf position for roughly 21,000 books~\cite{Broadbent2020}.
Books at eye level recorded about 2.3 times the in-library uses of books on the bottom shelf, and bottom-shelf items were used least on both checkouts and in-library use.
He attributes the pattern to physical browsing, while noting that the data cannot separate which of five possible retrieval processes generated any individual use.
None of it transfers to ebooks without new evidence.
A face-out physical display alters spatial exposure in ways a digital listing may not.

Finally, the user-level literature warns against reading non-use as self-explanatory, even for people.
Brunskill and Hanneke screened 1,864 records for their scoping review of academic-library non-users and included 69~\cite{Brunskill2021}.
Definitions of non-user varied widely, and 88 percent of the included studies relied on survey instruments.
Potnis et al.\ model intention to use ebooks among 279 undergraduates~\cite{Potnis2018}.
Casselden and Pears document search and reading pathways among 92 students, finding targeted interrogation within ebooks and a bite-size mode of engagement~\cite{Casselden2020}.
Both are informative about format-level adoption.
Neither reveals which individual titles were ever in any reader's awareness set.
Retrospective self-report cannot recover an exposure that went unnoticed.

%%%%%%%%%%%%%%%%%%%%%%%%%%%%%%%%%%%%%%%%%%%%%%%%%%%%%%%%%%%%%%%%%%%%%%%%%%%%%%%%
\section{THE FORMAL STATEMENT OF THE PROBLEM}

Recommender-system research met the identical structure years ago, and has a precise vocabulary for it.

Saito et al.\ treat implicit-feedback datasets as missing not at random~\cite{Saito2020}.
Whether an interaction is observed depends on whether the system exposed the item, so they derive estimators that correct for the exposure process rather than assuming it away.
Castells and Moffat state the consequence for evaluation directly~\cite{Castells2022}.
Offline metrics assume observed feedback proxies the whole preference space.
Feedback is missing not at random, because the system controls exposure.
Apparent non-preference is therefore, often, non-exposure.
That is the cleanest available statement of why ebook zero-use is not interpretable as demand.

Two further constructs transfer.
Popularity bias describes systems that amplify already-visible items at the expense of the long tail, unequally across user groups~\cite{Abdollahpouri2021}, with reinforcing feedback loops that survey work calls a Matthew effect~\cite{Klimashevskaia2024}.
Item cold start names a newly added item with no interaction history, structurally invisible to collaborative filtering until some independent route discovers it~\cite{Panda2022}.
That is an exact description of a newly acquired ebook.
Diaz et al.\ supply a rigorous definition of exposure as a quantity distinct from use, and an evaluation framework built on it~\cite{Diaz2020}.

The transfer must be flagged, not assumed.
These results come from commercial datasets in media and e-commerce.
None has been validated on an academic ebook collection.
The analogy is strong at the level of observational logic and unestablished at the level of scholarly reading behaviour.
What it does establish is this: the identification problem is real, formally characterised, and not resolvable by collecting more of the same usage data.

%%%%%%%%%%%%%%%%%%%%%%%%%%%%%%%%%%%%%%%%%%%%%%%%%%%%%%%%%%%%%%%%%%%%%%%%%%%%%%%%
\section{HOW THE PROBLEM HAS BEEN OPERATIONALISED}

Table~\ref{tab:ops} sets out the dependent variables in use, what each detects, and where each fails as a measure of encounter.
The ordering is informative.
Observability runs availability, exposure, selection, use, citation.
Every established library measure begins at the third stage or later.

\begin{table*}[!t]
\caption{Dependent variables in the 2018+ literature and their limits as measures of encounter}
\label{tab:ops}
\centering
\small
\begin{tabular}{@{}p{3.3cm}p{2.6cm}p{4.0cm}p{5.4cm}@{}}
\toprule
\textbf{Dependent variable} & \textbf{Exemplar evidence} & \textbf{What it detects} & \textbf{Principal validity threat} \\
\midrule
Binary title use / zero-use & \cite{Fry2018,Yilmaz2022,Strothmann2020} & At least one qualifying interaction in a window & Conflates non-exposure, unnoticed exposure, rejection, and off-channel use \\[3pt]
Raw item or page requests & \cite{Hughes2020,COUNTER51} & Intensity of interaction after content is reached & Delivery granularity changes totals; repeated requests are not repeated encounters \\[3pt]
Unique-title request & \cite{COUNTER51,Tracy2019} & Whether a session produced a qualifying title interaction & Improves the use denominator; creates no exposure denominator \\[3pt]
Link-resolver click & \cite{Echeverria2023} & Movement through a resolver toward provider content & Only moderately correlated with COUNTER; users who never click remain invisible \\[3pt]
Cross-platform download count & \cite{Snijder2021} & Platform-specific recorded access events & Different analytics systems yield materially different totals for one corpus \\[3pt]
Demand-driven trigger & \cite{Downey2020} & Demand crossing a vendor-defined threshold & Heavily selected endpoint; excludes titles noticed but not triggered \\[3pt]
Percent of accessible pool used & \cite{Yilmaz2022,Tran2021} & Fraction of an available pool generating use & Treats availability as the denominator when perceptual exposure is far smaller \\[3pt]
Citation count & \cite{Tyler2019} & Downstream scholarly attention & Very remote from first exposure; confounded by disciplinary citation practice \\[3pt]
Self-reported use or intention & \cite{Potnis2018,Casselden2020} & Attitude and adoption at format level & Silent on which individual titles entered awareness \\[3pt]
Library non-user status & \cite{Brunskill2021} & Person-level classification under heterogeneous definitions & Institution-level non-use cannot identify item-level non-exposure \\[3pt]
Encounter episode & \cite{Buchanan2019,Erdelez2020} & A contextually embedded incidental acquisition & Requires interview or diary reconstruction; does not scale to transaction logs \\[3pt]
Physical shelf position & \cite{Broadbent2020} & Effect of placement on print use & Print only; no validated digital analogue \\[3pt]
Physical display promoting ebooks & \cite{Tingle2018} & Whether a promoted title records any use in a window & Small $n$; multi-platform reporting gaps force unmeasured titles to count as unused \\[3pt]
Result or browse impression & \emph{None located} & Whether a title was rendered where it could be seen & Rendering is not noticing; requires attention measurement \\
\bottomrule
\end{tabular}
\end{table*}

Two consequences follow.
An observed zero identifies low value only if exposure is independent of item value and reader need, and the reviewed literature supplies no support for that assumption.
Newly acquired titles are the acute case.
An item that begins with no local usage history is exactly the item a use-driven ranking has no reason to surface.

%%%%%%%%%%%%%%%%%%%%%%%%%%%%%%%%%%%%%%%%%%%%%%%%%%%%%%%%%%%%%%%%%%%%%%%%%%%%%%%%
\section{DIRECTIONS THAT WOULD NARROW THE GAP}

We set out four measurement directions.
Each would resolve part of the identification problem.
None is a solution, and we state the residual in every case.

\subsection{Impression logging joined to requests}
Log, at title level, every occasion on which a record is rendered in a result list, carousel, or browse strip, and join those impressions to subsequent requests.
That supplies the exposure denominator the literature lacks.
It permits direct estimation of the conditional cascade
\begin{equation}
\begin{split}
P(\text{rendered}) &\to P(\text{noticed} \mid \text{rendered}) \\
                   &\to P(\text{selected} \mid \text{noticed}) \\
                   &\to P(\text{used} \mid \text{selected})
\end{split}
\label{eq:cascade}
\end{equation}
in place of a single undifferentiated usage rate.
The four terms match the four states of Section~IV, one for one.
\emph{Residual:} rendering is not noticing.
Without attention measurement an impression stays an upper bound on opportunity, and the second term in~(\ref{eq:cascade}) stays unidentified.

\subsection{A controlled test of a digital new-arrivals surface}
The shelf-position effect on print is documented~\cite{Broadbent2020}, and one study has already carried a display across to ebooks.
Tingle and Teeter printed cover images and short blurbs for 53 ebooks and shelved them by call number among the print stacks of a business library~\cite{Tingle2018}.
Seven of the 53 recorded use, a rate of 13.2 percent.
A stand-alone display at the same library had reached 12.3 percent, and estimated use across the business ebook collection that year was 10.82 percent.
The authors decline to claim an effect, judging a gap of 0.9 percentage points too small to be certain of.
This is an underpowered study rather than a null one, and it is instructive for a second reason: two of the eleven platforms involved returned no title-level data, so titles on those platforms were counted as unused by default.
The digital case remains open.
Clamon describes an automated new-books carousel driven by library-system analytics, with click tracking routed through the link resolver, and reports no measurement of any resulting change in use; formal evaluation is listed as future work~\cite{Clamon2019}.
Tran and Guo observe download peaks that coincide with promotional email to liaison departments, but with no counterfactual~\cite{Tran2021}.
A before-and-after or randomised assignment of newly acquired ebooks to a digital new-arrivals surface therefore remains untested, and inexpensive.
\emph{Residual:} a positive result would establish that one particular surface generates use.
It would not establish that the print shelf effect transfers, since face-out physical display alters exposure in ways a listing may not.
Tingle and Teeter also show what such a study must budget for: without title-level reporting from every platform in scope, the outcome measure is censored before the experiment begins.

\subsection{Including ebooks in call-number browse}
At least one major virtual-shelf implementation excludes electronic holdings by design~\cite{OCLC2024}.
Adding them and measuring the difference is a direct test of whether classification adjacency generates title-level exposure in a born-digital collection.
\emph{Residual:} a click on a virtual shelf may itself be goal-directed rather than incidental.
Bj\"orneborn's argument implies that adjacency without sensoriability may not reproduce the physical effect~\cite{Bjorneborn2017}.

\subsection{Partitioning zero-use by opportunity}
With an exposure log in hand, the zero-use population splits into never-rendered, rendered-and-unselected, and used-off-channel components.
Three interpretable categories replace one that is not.
\emph{Residual:} this needs exposure and use instrumented jointly, on the same collection, over the same window.
No study located under our protocol has done it, and off-channel use stays partly unobservable by construction.

None of these measures an encounter in the sense Erdelez and Makri define~\cite{Erdelez2020}.
They measure opportunity, which is the necessary condition.
Closing the rest of the distance needs attention data that libraries do not collect, and may have good reason not to.

%%%%%%%%%%%%%%%%%%%%%%%%%%%%%%%%%%%%%%%%%%%%%%%%%%%%%%%%%%%%%%%%%%%%%%%%%%%%%%%%
\section{LIMITATIONS}

Two kinds of limitation apply: to the evidence base, and to our own review.

The evidence base is skewed.
Geographically it concentrates in Anglophone research universities, with the Turkish and Spanish cases as partial exceptions.
We located no study set in a Netherlands or wider European context.
Sectorally it clusters in large research libraries, and the exposure evidence from recommender systems comes from commercial platforms.
By discipline, item-level ebook studies over-sample science, technology, engineering, and mathematics (STEM) collections, and interdisciplinary ones.
Humanities and social-science monograph behaviour, where encounter is valued most, is comparatively under-instrumented.
Several of the most directly relevant artefacts are grey literature or vendor documentation, so the material closest to the question is also the least evidentially robust.
The one located study of a display promoting ebooks is small and its outcome measure incomplete~\cite{Tingle2018}, so it neither supports nor rules out a display effect.
The proportions in Table~\ref{tab:zerouse} are not comparable across rows: windows, platforms, subject mixes, entitlement models, and metrics all differ between studies.

Our own review is limited in the ways set out in Section~III-D.
Two authors conducted it, without double-screening.
It retained no stage-by-stage record counts, and it is not reproducible in the sense a systematic review aims at.
The negative claims are therefore claims about what a structured search did not surface.
We regard them as strong enough to motivate the measurement directions of Section~IX, and not strong enough to close the question.
A registered systematic review of exposure measurement in born-digital library collections would be a worthwhile successor to this paper.

%%%%%%%%%%%%%%%%%%%%%%%%%%%%%%%%%%%%%%%%%%%%%%%%%%%%%%%%%%%%%%%%%%%%%%%%%%%%%%%%
\section{CONCLUDING REMARKS}

The literature published since 2018 measures what happens after a reader acts on a title, and measures it with considerable precision.
What happens before that act it barely measures at all.
Large zero-use populations are firmly established.
Their interpretation is not.
The strongest defensible conclusion is narrower and more uncomfortable than the one usually drawn: standard usage evidence cannot tell whether an unused ebook was ever meaningfully considered.

The operational consequence is direct.
Assessment that reads recorded non-use as revealed preference will systematically disadvantage the very titles the system failed to surface.
Where ranking is use-driven, the disadvantage compounds.
The category most exposed is the newly acquired title, which begins with no history and therefore no reason to be shown.

An ebook can be technically findable while its existence is unknown to every reader who would have valued it.
In what sense has it been discovered?
The literature has strong instrumentation for the moment after the reader acts.
For the moment in which an unknown book first becomes a possible book, it has very little.

%%%%%%%%%%%%%%%%%%%%%%%%%%%%%%%%%%%%%%%%%%%%%%%%%%%%%%%%%%%%%%%%%%%%%%%%%%%%%%%%
\section*{ACKNOWLEDGMENT}

% EDIT: name the specific colleagues who contributed, or remove this section.
The authors thank [NAMES] for discussion of the collection-assessment practices that motivated this review.

%%%%%%%%%%%%%%%%%%%%%%%%%%%%%%%%%%%%%%%%%%%%%%%%%%%%%%%%%%%%%%%%%%%%%%%%%%%%%%%%

\printbibliography

%\addtolength{\textheight}{-2cm}    % Uncomment and tune to balance the column
                                    % lengths on the last page of the document

\end{document}